\documentclass[%
reprint,
superscriptaddress,
amsmath,amssymb,
aps,
prl,
]{revtex4-1}
\usepackage{graphicx}
\usepackage{dcolumn}
\usepackage{bm}
\usepackage{amsmath}
\usepackage{hyperref}
\hypersetup{colorlinks=true,linkcolor=blue,anchorcolor=blue,citecolor=blue,urlcolor=magenta}
\usepackage{lineno}
\usepackage{enumerate}
\usepackage{color}

\begin{document}
\preprint{APS/123-QED}

\title{Anisotropy and energy distribution of leptons and photons in radiative relativistic Alfv\'{e}nic turbulence}

\author{Peng Liu}
\affiliation{Institute of Theoretical Physics, Chinese Academy of Sciences, Beijing 100190, China}

\author{Xiaochun Wei}
\affiliation{Institute of Theoretical Physics, Chinese Academy of Sciences, Beijing 100190, China}

\author{Zheng Gong}
\email{zgong92@itp.ac.cn}
\affiliation{Institute of Theoretical Physics, Chinese Academy of Sciences, Beijing 100190, China}%

\date{\today}

\begin{abstract}

The anisotropic evolution of relativistic particles in magnetically dominated Alfv\'{e}nic turbulence is a fundamental process governing nonthermal emissions in high-energy astrophysical environments. 
In this work, we develop a theoretical framework by deriving scaling equations that characterize the pitch-angle dynamics of leptons and photons across both weak- and strong-radiative cooling regimes. 
Our model reveals a universal coupling between particle energy and angular distribution, arising from the interplay of field-aligned accelerations, perpendicular drift motions, and radiation reactions. The model can uniquely identify ``turning-point energies" that exhibit a nonlinear dependence on the local magnetic field and the energy-injection scale. The validity of this scaling model is confirmed through radiative particle-in-cell simulations.
We show that the electron pitch-angle distribution fundamentally dictates photon emission characteristics, further revealing a pronounced statistical correlation between high-energy photon emission and turbulent vortices in the reconnection regions.
By bridging the gap between small-scale kinetic interactions and broad-band spectral characteristics, this work provides a quantitative basis for interpreting the hardening of radiation spectra and complex polarization signatures in sources such as pulsar wind nebulae and gamma-ray bursts.

\end{abstract}
\maketitle
Plasma turbulence is a quintessential multi-scale and chaotic phenomenon, ubiquitous in magnetically dominated astrophysical environments~\cite{Hillebrandt2009book,Howes2015book}. 
While energy transfer from macro-scales to kinetic scales is well-described by magnetohydrodynamic (MHD) frameworks~\cite{Kraichnan1965MHD,Schekochihin2009,Matthaeus2011MHD, Loureiro2017MHD,Schekochihin2022,Dong2022MHD,Giuseppe2025MHD}, the dissipation mechanisms at scales below the ion gyroradius necessitate a kinetic treatment~\cite{Howes2008kinetic,Howes2015book,Groselj2019kinetic,Cerri2021kinetic,Arzamasskiy2023kinetic,Park2025kinetic}.
Unlike the strongly collisional limit of MHD, collisionless kinetic turbulence enables energy dissipation via wave-particle interactions, with prominent examples including kinetic Alfv\'en turbulence~\cite{Peter2009KAW,Salem2012KAW,Boldyrev2012KAW,Daniel2018KAW, Zhou2023KAW,David2024SolarKinetic} and Weibel-instability-driven turbulence~\cite{Sheng2015WeibelTurb,Sironi2023WeibelTurb,Liu2024WeibelTurb,Zhang2024WeibelTurb,Yuan2024WeibelTurb}. 
Crucially, first-principles particle-in-cell (PIC) simulations have revealed that nonthermal particle acceleration is an intrinsic feature of kinetic turbulence, whether in the large-amplitude turbulence regime~\cite{Zhdankin2017,Zhdankin2018,Zhdankin2019,ComissoPRL2018,ComissoAPJ2019,ComissoAPJL2020,Comisso2021,Sobacchi2021, ComissoAPJL2022,Vega2022,Bresci2022,Zhang2023,Imbrogno2024,Gorbunov2025,Lemoine2025,Zhu2025} or the weakly excited Alfv\'{e}nic turbulence (WEAT) limit~\cite{ComissoPRL2018,Trotta2020,Nattila2022,Vega2024,Vega2025}.

In relativistic turbulence systems such as pulsar wind nebulae (PWNe)~\cite{Buhler_2014,Kargaltsev2015,Xie2022}, active galactic nuclei (AGNs)~\cite{Marscher2008AGN, Madejski2016,Bourne2017}, and gamma-ray bursts (GRBs)~\cite{Piran2005,Meszaros2006,Narayan2009,Burgess2020}, kinetic effects are fundamentally coupled with photon emission~\cite{Comisso2021,ComissoAPJL2020,Nattila2021,Groselj2024,Nattila2024radiative} and radiative cooling~\cite{Uzdensky2018,Zhdankin2020,Zhdankin2021,Sobacchi2021,Comisso2021,Bacchini2024}. 
When turbulence reaches the quantum electrodynamic regime~\cite{Nattila2024radiative,Mehlhaff2025,Liu2026arxiv}, discrete photon emission becomes a dominant energy loss channel. In these environments, individual leptons can transfer a significant fraction of their energy to single photons, reaching X-ray or even gamma-ray energies~\cite{Meszaros2006,Buhler_2014,Madejski2016}. This radiative feedback enriches the kinetic physics, requiring precise tracking of particle trajectories and their associated radiative losses.

\begin{figure*}[t]
  \includegraphics[scale=0.41]{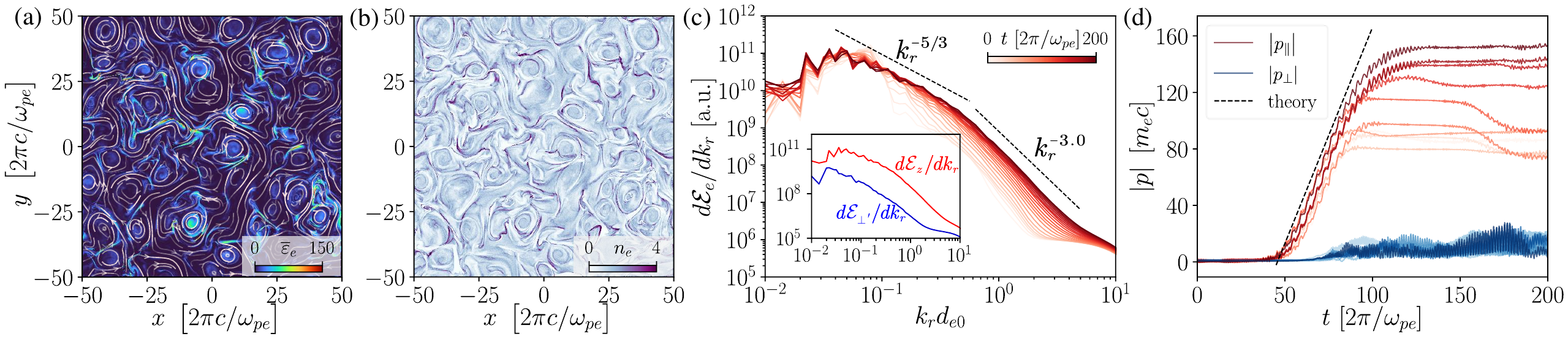}
  \caption{\label{fig:figure1} Simulation results with a magnetization of $\sigma_{\delta B} = 50$. Time snapshots of spatial distributions of (a) averaged electron kinetic energy $\overline{\varepsilon}_e$ normalized to $m_e c^2$ and (b) electron density $n_e$ normalized to $n_0$ at $t = 400\pi/\omega_{pe}$, where the streamlines in (a) denote the direction of the magnetic field $\delta \mathbf{B}_{xy}$.
  (c) Time evolution of the kinetic energy spectrum of electron velocity flows $d\mathcal{E}_e/dk_r$. The inset shows the separated energy spectra of the $z$-aligned component $d\mathcal{E}_z/dk_r$ and the in-plane component $d\mathcal{E}_{\perp'}/dk_r$ at $t = 400\pi/\omega_{pe}$.
  (d) Time evolution of the momenta $p_\parallel$ and $p_\perp$ for 10 representative high-energy electrons, where the evolution of $p_\parallel$ is approximated by $ p_{\parallel}  \simeq 3.1 t - p_{\parallel 0}$~\cite{ComissoPRL2018,ComissoAPJ2019} with $p_{\parallel 0}\simeq 150 m_ec$ denoted as the black dashed line.}
\end{figure*}

Recent simulations have shown that leptons are initially energized by reconnection-induced electric fields within turbulent current sheets~\cite{ComissoPRL2018,ComissoAPJ2019,Comisso2021,Nattila2021,Singh2025}, followed by stochastic (Fermi-like) acceleration~\cite{Fermi1949,Fermi1954,Tramacere2011,Petrosian2012,Blasi2013,Lemoine2022,Kempski2023,Kundu2023,ComissoPRL2018, ComissoAPJ2019,Comisso2024,Lubke2025}. 
A hallmark of this energization is the progressive deviation from pitch-angle isotropy~\cite{ComissoAPJ2019,ComissoAPJL2020,Comisso2021,ComissoAPJL2022}. While synchrotron emission is suppressed at small pitch angles, radiative cooling further regulates the angular distribution~\cite{Comisso2021}. 
Despite its importance, the physical mechanism governing this energy-dependent anisotropy and its interplay with strong radiative cooling remains a critical, unresolved challenge~\cite{Nattila2022,Comisso2021,ComissoAPJL2020}. 
Establishing a quantitative model for pitch-angle evolution is therefore paramount, as it dictates whether radiative power can exceed the nominal radiation-reaction limit~\cite{Comisso2021}. Such a model is essential for interpreting astrophysical phenomena, including the hard radio spectra of PWNe~\cite{Lyutikov2019}, as well as the flares with fast rise and decay times in X-ray afterglow of GRBs~\cite{Beloborodov2011} and the counterintuitive high-degree circular polarization in optical afterglow of GRBs~\cite{Wiersema2014,Nava2015}.

In this article, we establish an analytical scaling model for the energy-dependent pitch-angle anisotropy of both leptons and photons in WEAT. Our model, validated by $ab\ initio$ radiative PIC simulations, captures the universal scaling laws across diverse energy regimes. 
The model shows that the observed hardening of the electron energy spectrum and the flattening of the pitch-angle distribution are direct consequences of intensified radiative cooling in WEAT, serving as a complement to the framework of large-amplitude turbulence~\cite{Comisso2021}. 
By employing detailed particle tracking, we further reveal a pronounced statistical correlation between high-energy emission and turbulent vortices in reconnection regions. 
These findings potentially offer new insights into the anomalous fast variability, high-degree circular polarization, and unexpected hard-radio spectra observed in high-energy astrophysical environments~\cite{Wiersema2014,Nava2015,Beloborodov2011}.

We perform 2D simulations of decaying Alfvénic turbulence using the radiative particle-in-cell (PIC) code EPOCH~\cite{Arber2015}. 
The initialized plasma consists of electron-positron pairs, following an isotropic Maxwell-J\"uttner distribution with a normalized temperature $\Theta_0 \equiv T_0/m_ec^2=0.3$. 
The simulation domain is a square box of size $L^2=(100\lambda_{e0})^2$, where $\lambda_{e0}=2\pi c/\omega_{pe}$ is the plasma skin depth, $\omega _{pe}=\sqrt{4\pi n_0e^2/m_e}$ the plasma electron frequency, $e$ ($m_e$) the electron charge (mass), and $n_0$ the density.
The spatial resolution is set to $\Delta x = \Delta y = \lambda_{e0}/40$ to resolve the characteristic kinetic scales adequately. 

The system is initially permeated by a local magnetic field $\mathbf{B}_{\mathrm{loc}}=B_0\mathbf{\widehat{z}}+\delta \mathbf{B}_{xy}$ with $\delta \mathbf{B}_{xy}=\left( \delta B_{x0},\delta B_{y0} \right) $.
To drive the decaying turbulence, we initialize magnetic fluctuations $\delta \mathbf{B}_{xy}$ following established protocols~\cite{ComissoPRL2018,ComissoAPJ2019,ComissoAPJL2020,ComissoAPJL2022,Vega2022,Vega2024,Vega2025,Comisso2021,Nattila2021}, with $\delta B_{x0}=\sum_{m,n}{\mathcal{A}_{mn}n\sin \left( k_mx+\phi _{mn} \right) \cos \left( k_ny+\varphi _{mn} \right)}$ and $\delta B_{y0}=-\sum_{m,n}{\mathcal{A}_{mn}m\cos \left( k_mx+\phi _{mn} \right) \sin \left( k_ny+\varphi _{mn} \right)}$, where $m\left( n \right) \in \left[ 1,...,N_{\max} \right] $ indicate the mode number in $x\ (y)$ direction with $N_{\max}=8$, $k_{m\left( n \right)}=2\pi m\left( n \right) /L$ is wave number, $\phi_{mn}$ and $\varphi_{mn}$ are random phases, $\mathcal{A}_{mn}=2\delta B_{\mathrm{rms}0}/N_{\max}\left( m^2+n^2 \right) ^{1/2} $ with $\delta B_{\mathrm{rms}0}=\left< \delta B_{x0}^{2}+\delta B_{y0}^{2} \right> ^{1/2} $ being the root-mean-square fluctuating magnetic field at $t=0$, and $\left<...\right>$ denotes averages.
For the WEAT regime studied here, the initial root-mean-square fluctuation is fixed at $\delta B_{\mathrm{rms}0}/B_0=1/10$. 
The primary physical parameter varied across our simulations is the relativistic hot magnetization,
$\sigma_{\delta B} =\delta B_{\mathrm{rms}0}^{2}/4\pi n_0h_0m_ec^2 $, where $h_0=4\Theta _0+K_1\left( 1/\Theta _0 \right) /K_2\left( 1/\Theta _0 \right) \simeq 1.9 $ is the initial enthalpy per particle. $K_1$ and $K_2$ are the modified Bessel functions of the second kind with orders 1 and 2. 
The $\sigma_B = \sigma_{\delta B} B_0^2/ \delta B_{\mathrm{rms}0}^2$ is also defined.
In our fiducial run, $\sigma_{\delta B}=50$. To ensure the universality of our scaling laws, we further explore a broad parameter space with $ \sigma_{\delta B} \in \{2,4.5,8,18,32,112.5 \}$ and system sizes of $L/\lambda_{e0} \in \{25,50,200,300 \} $~\cite{Supplement}.

To elucidate the lepton dynamics in the WEAT regime with $\sigma_{\delta B} =50$, we present the spatial distribution of average electron kinetic energy and density in Figs.\ \ref{fig:figure1}(a) and (b) \footnote{As positrons and electrons would exhibit analogous behavior, here we focus on electron characteristics for brevity}. 
The energized particles exhibit high spatial localization, primarily concentrated within the vicinity of current sheets, with electrons accumulating around turbulent vortices. This localization stands in stark contrast to large-amplitude turbulence ($\delta B_{\mathrm{rms}0}/B_0 \sim 1$), where energetic particles are outside the current sheets and more uniformly dispersed throughout the entire simulation domain ~\cite{ComissoAPJ2019,Nattila2021}. 
The time evolution of the kinetic power spectrum, $d\mathcal{E}_e/dk_r=\left| \text{fft}\left( \sqrt{n_e}\overline{\mathbf{v}} \right) \right|^2$ with $\overline{\mathbf{v}}$ being the averaged electron velocity, is shown in Fig.\ \ref{fig:figure1}(c). The spectrum displays a characteristic broken power-law decay, following $\sim k_r^{-5/3}$ at large scales ($k_{r} \sim 0.1d_{e0}^{-1} $) and steepening to $\sim k_r^{-3.0}$ at kinetic scales $k_{r} \sim d_{e0}^{-1}$, where $k_{r} = \left( k_x^2+k_y^2 \right)^{1/2}$. 
Remarkably, as illustrated in the insets, the kinetic power is almost entirely dominated by the $z$-aligned velocity component. This indicates that the particle energization is driven predominantly by the work done by the local parallel electric field, $ \mathbf{E}_{\parallel}=\left( \mathbf{E}\cdot \mathbf{B}_{\mathrm{loc}} \right) \mathbf{B}_{\mathrm{loc}}/B_{\mathrm{loc}}^2 $, a process further confirmed by our particle-tracking analysis. 
Consequently, energetic particles preferentially gain parallel momentum, $ p_\parallel $, whose evolution can be approximately modeled by the normalized equation $dp_{\parallel}/dt \simeq 2\pi E_{\parallel}/\delta B_{\mathrm{rms}0}\sqrt{\sigma_{\delta B}\left( \left< \gamma_0 \right> +\Theta_0 \right)}$. Assuming $\left| E_{\parallel}/\delta B_{\mathrm{rms}0} \right| \sim 0.05$, it yields a linear solution  $p_{\parallel}  \simeq 3.1 t - p_{\parallel 0}$~\cite{ComissoPRL2018,ComissoAPJ2019}, showing excellent agreement with our simulation results [see the dashed line in Fig.\ \ref{fig:figure1}(d)]. This parallel-dominant energy gain provides the physical origin for the pronounced momentum and pitch-angle anisotropy observed in the WEAT regime, which is consistent with the simulation results of the injection phase shown in Refs.~\cite{ComissoPRL2018,ComissoAPJ2019}

\begin{figure}[t] 
  \includegraphics[scale=0.45]{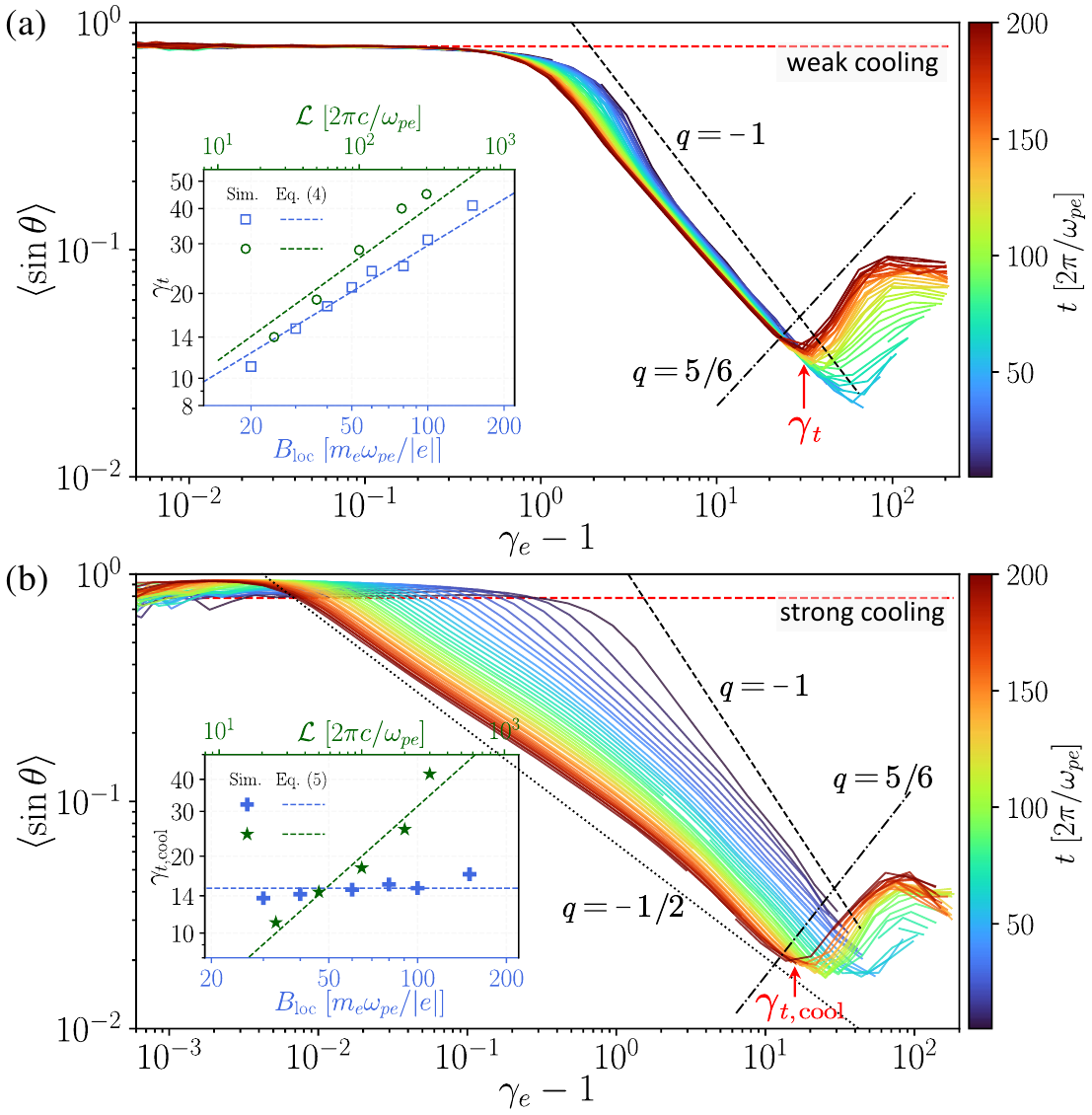}
  \caption{\label{fig:figure2} Time evolution of the electrons' mean pitch angle $\left< \sin \theta \right>$ in the weak (a) and strong (b) cooling cases with $\sigma_{\delta B} = 50$. The black dashed, dot-dashed, and dotted lines denote the predicted scaling laws of Eqs.~\eqref{equation2} ($q=-1$), \eqref{equation3} ($q=5/6$), and \eqref{equation5} ($q=-1/2$), respectively. The horizontal red line refers to the pitch angle  $\left< \sin \theta \right>=\pi/4$ for an isotropic particle distribution.
  The insets show the dependencies of the turning-point energies $\gamma_t$ and $\gamma_{t,\mathrm{cool}}$ on $B_{\mathrm{loc}}$ (blue) and $\mathcal{L}$ (green) at $t=400\pi/\omega_{pe}$, where the dashed lines denote the theoretical predictions of Eqs.\ (\ref{subequ6-1}) and (\ref{subequ6-2}).}
\end{figure}

To quantify the dependence of pitch-angle anisotropy on particle energy and radiative cooling, we examine the evolution of the pitch angle under varying radiative intensities. The cooling strength is characterized by the quantum strong-field invariant,
$\chi _{e} \equiv \left( e \hbar /m_{e}^{3}c^4 \right) \left| F_{\mu \nu}p^{\nu} \right| \simeq \gamma_e \left| \mathbf{B}_{\mathrm{loc}}\times \mathbf{p} \right|/\left| \mathbf{p} \right|B_S$, where $F_{\mu \nu}$ and $p^\nu$ are the electromagnetic field tensor and electron four-momentum, respectively, 
$B_S=m_{e}^{2}c^2/ e \hbar \simeq 4.4\times 10^{13}\ \mathrm{G}$ is the Schwinger field, $\hbar$ is the reduced Planck constant, and $\mathbf{p}$ is the electron momentum.
Figure \ref{fig:figure2}(a) shows the ensemble-averaged $\sin\theta$ as a function of the kinetic energy $(\gamma_e - 1)$ for a weak-cooling case with $B_{\mathrm{loc}} \simeq 10^{10}$ G and $\chi_{e,\max} \sim 4 \times 10^{-3}$. 
Following established methodologies~\cite{ComissoAPJ2019,ComissoAPJL2020,Comisso2021,Nattila2022,Vega2025,Saikat2025arxiv}, both $\sin\theta$ and $\gamma_e$ are measured in the local $\mathbf{E} \times \mathbf{B}$ frame. In this frame, the perpendicular electric field $\mathbf{E}_{\perp}$ vanishes \footnote{In the ideal MHD approximation, the total $\mathbf{E}$ can be vanished in the $\mathbf{E}\times \mathbf{B}$ frame~\cite{Bresci2022}, in which synchrotron radiation loss can be directly evaluated}, allowing for a direct evaluation of synchrotron losses while suppressing laboratory-frame energy oscillations associated with gyromotion~\cite{Comisso2021}.
For low-energy particles in this regime, our simulations reveal that the perpendicular momentum remains approximately constant, i.e. $p_\perp \simeq p_c$, relative to the rapidly increasing parallel momentum. Consequently, the pitch angle $\theta_{\gamma^<}$ can be analytically expressed as a function of the Lorentz factor $\gamma_e$:
\begin{equation}\label{equation2}
  \sin \theta _{\gamma ^{<}} \equiv \frac{p_{\perp}}{p}=  \mathcal{C} \left( \gamma_{e}^2-1 \right)^{-1/2},
\end{equation}
where $\mathcal{C} = p_c/m_ec$ is a constant. In the relativistic limit where $1 \ll \gamma_e \lesssim 30$, Eq.~(\ref{equation2}) yields the scaling law $\sin \theta_{\gamma^<} \propto \gamma_e^{-1}$, this behavior is inherently a natural consequence of electrons being accelerated by the reconnection electric field along the magnetic field lines, which has been discussed in the relativistic magnetic reconnection frameworks~\cite{ComissoMR2023,ComissoMR2024}.

As particles reach higher energies, their perpendicular dynamics are increasingly governed by drift motions [see Fig.\ \ref{fig:figure1}(d)]. 
In the relativistic limit where $v_{\parallel} \sim c \gg v_{\perp}$, the effective perpendicular momentum can be approximated by the curvature drift contribution: $p_{\perp} \simeq \gamma_e^{2} \rho_0 m_e c |\boldsymbol{\kappa}|$. Here, $\rho_0 = m_e c / |e| B_{\mathrm{loc}}$ is the characteristic gyroradius scale, and $\boldsymbol{\kappa} = \mathbf{\hat{b}} \cdot \nabla \mathbf{\hat{b}}$ denotes the magnetic field-line curvature with $\mathbf{\hat{b}} = \mathbf{B}_{\mathrm{loc}} / | \mathbf{B}_{\mathrm{loc}} |$.
In the WEAT regime, assuming $\partial_z \ll \nabla_{\perp^ \prime}$ where ${\perp^ \prime}$ refers to the direction transverse to the background field along $z$-direction, the curvature radius is estimated as $r_c=1/\left| \boldsymbol{\kappa } \right|\sim B_{0}^{2}l_{\perp^ \prime}/\delta B_{\mathrm{rms}}^2$. 
To link this with turbulence scales, we invoke the critical balance conjecture~\cite{Goldreich1995}, which posits that the Alfv\'{e}n time and nonlinear interaction time are comparable, i.e., $l_z/l_{\perp^ \prime}\sim B_0/\delta B_{\mathrm{rms}}$. 
Combined with the Kolmogorov phenomenology~\cite{Kolmogorov1941,Canuto2009}, one obtains $l_z\sim l_{\perp^ \prime}^{2/3}\mathcal{L}^{1/3} $~\cite{Goldreich1995,Thompson1998,Schekochihin2007,SchekochihinBook2007,Chen2016,Schekochihin2022}, where $l_{\perp^\prime}$ is the characteristic perpendicular scale of turbulence and $\mathcal{L}\ (\simeq L)$ is the energy-injecting scale. 
Consequently, the perpendicular momentum scales as $p_{\perp} \sim \gamma_e^{2}\rho_0 m_e c/l_{\perp^ \prime}^{1/3}\mathcal{L}^{2/3}$. 
Considering the characteristic energy dissipation scale with the relativistic inertial scale,  $l_{\perp^ \prime} \sim d_e=c/\sqrt{4\pi n_0e^2/\gamma_e m_e}$, we derive the analytical scaling for the high-energy pitch angle $\theta _{\gamma ^{>}}:$
\begin{equation}\label{equation3}
  \sin \theta _{\gamma ^{>}}  \propto \gamma_{e}^{5/6}\frac{\rho _0}{\mathcal{L}^{2/3}d_{e0}^{1/3}},
\end{equation}
where $d_{e}^{}=\sqrt{\gamma_e} d_{e0}$ and $p \simeq \gamma_e m_ec $.
These predictions Eqs.\ (\ref{equation2}) and (\ref{equation3}) are compared with the angular distribution of $\sim 6\times 10^6$ subsample particles in our weak-cooling simulations [Fig.\ \ref{fig:figure2}(a)]. The excellent agreement (black dashed and dot-dashed lines) validates our model.
Notably, even at the highest energies, $\langle \sin \theta \rangle$ remains significantly below the isotropic threshold of $\pi/4$ shown by red dashed line.
This stands in distinct contrast to large-amplitude turbulence with $\delta B_{\mathrm{rms}0} \gtrsim B_0$, where high-energy particles often exceed $\langle \sin \theta \rangle = \pi/4$~\cite{ComissoAPJL2020,Comisso2021}. We also note that at the thermal limit ($\gamma_e \to 1$), the distribution remains perfectly isotropic, as these particles have not yet undergone the anisotropic acceleration process.

As particle energies approach $\gamma_e \sim 100$, we observe an increasing divergence between the simulation results and the weak-cooling theoretical predictions. This discrepancy arises as radiative cooling becomes a dominant factor in the particle dynamics.
In order to investigate this, we present the simulation result of the energy-dependent pitch-angle distribution for a strong-cooling regime with $\chi_{e,\max}\sim 0.2$ and $B_{\mathrm{loc}}\simeq 10^{12}\ \mathrm{G}$ \footnote{It indicates that $e^{-}e^{+}$ pair production through the linear and nonlinear Breit-Wheeler process is negligible~\cite{Di2012}.} in Fig.\ \ref{fig:figure2}(b). In this regime, the distribution $f\left( \gamma _e-1,\left< \sin \theta \right> \right) $ shifts towards lower energies over time as a direct consequence of enhanced radiative losses.
A typical feature in the strong-cooling case is the evolution of the scaling index $q=d\log \left< \sin \theta \right> /d\log \left( \gamma _e-1 \right) $. For low-energy particles, $q$ gradually decreases and eventually converges to $\sim-1/2$ as the system cools [see black dotted line in Fig.\ \ref{fig:figure2}(b)].
By considering single-particle dynamics and incorporating radiation reaction, we developed a model to characterize the relationship between particle energy and pitch angle. 
To characterize this relationship, we incorporate the radiation reaction force into the single-particle equations of motion, $ d\mathbf{p}/dt=-\left| e \right|\mathbf{v}\times \mathbf{B}_{\mathrm{loc}}-\eta \chi _{e}^{2}\boldsymbol{\beta} $, 
where $\eta =4\pi\alpha_f m_ec^2 /3\lambda_C$, $\boldsymbol{\beta}=\mathbf{v}/c$, $\alpha_f$ is the fine-structure constant, and $\lambda_{C}$ is the Compton wavelength. 
Approximating the invariant as $\chi_e\simeq \gamma_e B_{\mathrm{loc}}\sin \theta /B_S\simeq p_{\perp}B_{\text{loc}}/m_ecB_S$, the evolution of the perpendicular momentum is governed by $dp_{\perp}/dt = -p_{\perp}^{3}\eta B_{\mathrm{loc}}^{2}/\gamma_em_{e}^{3}c^3B_{S}^{2}$~\cite{Supplement}.
Assuming that $\gamma_e$ changes slowly for low-energy electrons due to balanced radiative losses, we derive the scaling law governed by radiative cooling: 
\begin{equation}\label{equation5}
  \sin \theta_{\gamma_{\mathrm{cool}}} \simeq \left( \gamma_{e}-1 \right)^{-1/2}\frac{B_S}{2B_{\mathrm{loc}}}\sqrt{\frac{m_ec}{\eta t}}.
\end{equation}
This analytical result is consistent with the simulation results, where the slope exhibits clear time-dependence and asymptotically evolves to $q_{\min} = -1/2$ by $t=400\pi/\omega_{pe}$. 
Physically, the pitch angle is determined by the dynamic interplay between radiative cooling and perpendicular drift. 
For the majority of the low-energy population and the high-energy tail, radiative cooling exerts a more dominant influence on the lateral momentum than curvature drift, leading to the observed deviation from $q=-1$ and $5/6$ scalings. 
In contrast, perpendicular drift remains the primary influence for the intermediate-energy population ($20 \lesssim \gamma_e \lesssim 80 $).

The scaling laws derived above [Eqs.\ (\ref{equation2})-(\ref{equation5})] are robust across varying magnetic field strengths $B_{\mathrm{loc}}$ and energy-injection scales $\mathcal{L}$.
They can provide a quantitative explanation for the energy-dependent pitch-angle contraction in the nonthermal scenarios with $\delta B_{\mathrm{rms}0} \ll B_0 $ [Fig.~\ref{fig:figure2}]. Furthermore, our framework could be generalized to characterize turbulence in the large-amplitude regime ($\delta B_{\mathrm{rms}0} \sim B_0 $)~\cite{ComissoAPJ2019,ComissoAPJL2020,Comisso2021,ComissoAPJL2022,Zhu2025} (see the SM for a detailed discussion~\cite{Supplement}), and it may inspire future extensions to quasithermal or log-normal distribution regimes~\cite{Nattila2022,Vega2025}.

A fundamental implication of our model is the existence of a minimum in the pitch-angle distribution, where the governing dynamics transition between drift and acceleration or cooling.
By equating $\sin \theta _{\gamma ^>}$ with $\sin \theta _{\gamma ^<}$ and $\sin \theta_{\gamma_{\mathrm{cool}}}$, we derive the characteristic turning-point Lorentz factors for the weak- and strong-cooling regimes, respectively:
\begin{equation}\label{subequ6-1}
  \gamma_{t} \propto B_{\mathrm{loc}}^{6/11} \mathcal{L}^{4/11}d_{e0}^{2/11} \mathcal{C}^{6/11} \left( \frac{\left| e \right| }{m_{e}c} \right) ^{6/11}
\end{equation}
and
\begin{equation}\label{subequ6-2}
  \gamma_{t,\mathrm{cool}} \propto \mathcal{L}^{1/2}d_{e0}^{1/4} \left( \frac{e^2B_{S}^{2}}{2m_ec\eta t} \right) ^{3/8}.
\end{equation}
To verify these predictive scalings, we extract the turning-point electron energies from our simulations across a wide parameter space of $B_{\mathrm{loc}}$ and $\mathcal{L}$. As shown in the insets of Fig.\ \ref{fig:figure2}, the simulation values (markers) are in excellent agreement with the theoretical predictions (blue and green dashed lines). 
Notably, while $\gamma_t$ in the weak-cooling regime is jointly determined by the magnetic field and system size, $\gamma_{t,\mathrm{cool}}$ in the strong-cooling scenario exhibits a distinct dependence, becoming decoupled from the local field strength and governed primarily by the energy-injection scale $\mathcal{L}$.

\begin{figure}[t]
  \includegraphics[scale=0.37]{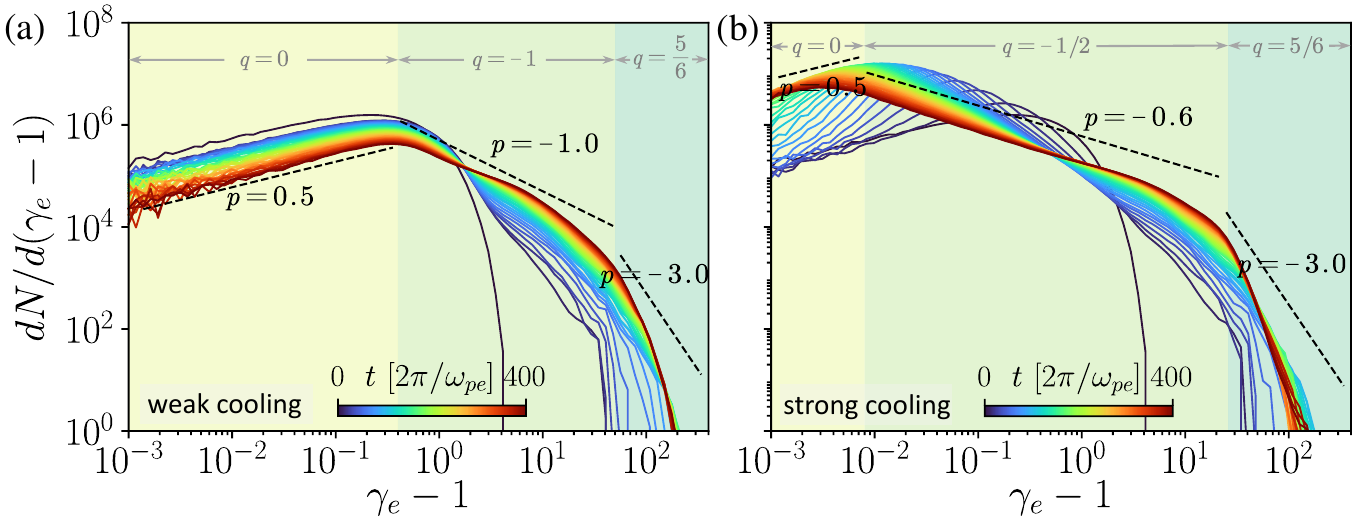}
  \caption{\label{fig:figure3} Time evolution of the electron energy spectra for the weak (a) and strong (b) cooling turbulence with $\sigma_{\delta B} = 50$. The three shaded regions correspond to the ranges with different pitch-angle scalings of $q$ in Fig.~\ref{fig:figure2}. The dashed lines denote the power-law indices $p$ in the corresponding $q$ ranges. Here, the power scalings of $p$ agree reasonably well with the predictions of Eq.~(\ref{empirical}).}
\end{figure}
A counterintuitive finding from our simulations is that radiative cooling manifests more significantly in the moderate-energy population than in the high-energy tail. This is evidenced by the evolution of the electron energy spectra [Fig. \ref{fig:figure3}], in which as cooling is intensified, the moderate-energy spectrum undergoes pronounced hardening, with the power-law index $p=d\log \left[ dN/d\left( \gamma_e-1\right) \right] /d\log \left( \gamma_e-1\right) \simeq -1 $ to $-0.6$. Note that this spectral hardening behavior is highly similar to the intense cooling regime investigated in the previous work with $\delta B_{\mathrm{rms}0} = B_0$~\cite{Comisso2021}, where the power-law index shifts from $p \sim -1.2$ to $p \sim -0.5$.
In contrast, the lower energy ($q\simeq0$) and high-energy ($q\simeq5/6$) regimes maintain remarkably consistent slopes with $p\simeq 0.5$ and $p \simeq -3$ across both weak- and strong-cooling scenarios. 
To reconcile these observations, we propose a model that couples the spectral index $p$ to the pitch-angle scaling index $q$:
\begin{equation}\label{empirical}
p  \simeq
\begin{cases} 
 q^2+2.5q+0.5 & \text{for }\ \ -1 \leq q\leq 0 \\ 
 q-4      & \text{for }\ \  0<q \leq 1
\end{cases}.
\end{equation}
This relationship effectively captures the spectral evolution in the WEAT regime and remains applicable to large-amplitude turbulence (see SM~\cite{Supplement} for verification).
For the high-energy tail  $\gamma_e \gtrsim 100$, the simulated spectral slope is slightly steeper than the prediction of Eq.\ (\ref{empirical}). This discrepancy stems from the omission of radiative corrections in the high-energy limit of our analytical model.

\begin{figure}[t] 
  \includegraphics[scale=0.41]{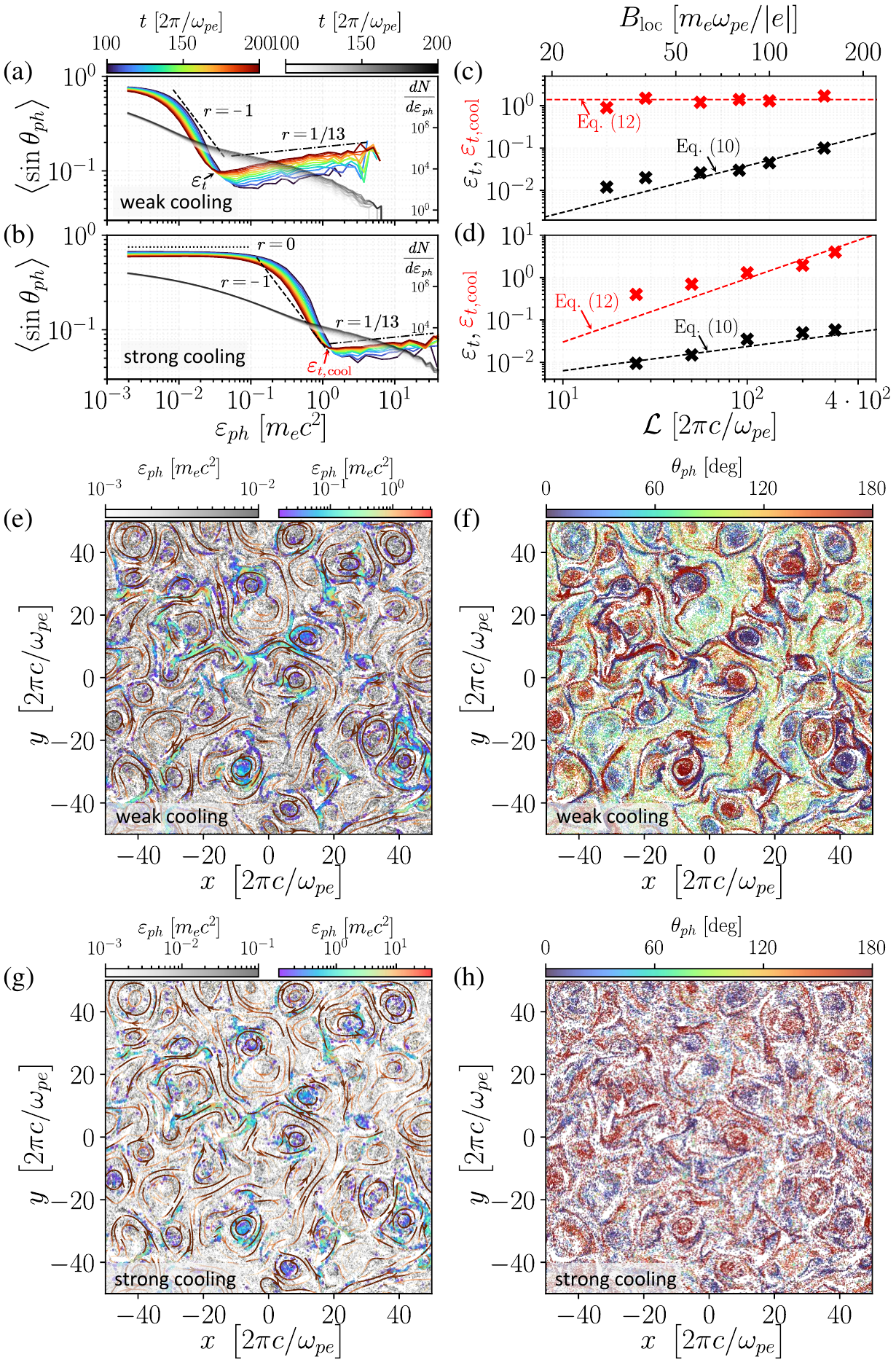}
  \caption{\label{fig:figure4} Time evolution of the mean pitch angle $\left<sin\theta_{ph}\right>$ and energy spectra of photons for the weak (a) and strong (b) cooling regimes, where the black dashed and dot-dashed lines show the predicted scalings determined by Eqs.\ (\ref{equation7}) and (\ref{equation9}).
  [(c),(d)] Dependencies of $\varepsilon_t$ (black) and $\varepsilon_{t,\mathrm{cool}}$ (red) on $B_{\mathrm{loc}}$ and $\mathcal{L}$.
  Spatial distribution of photon energy with different energy intervals in the weak (e) and strong (g) cooling scenarios, where the streamlines denote the magnetic field direction of $\delta \mathbf{B}_{xy}$.
  Spatial pitch-angle distributions of photons in the weak (f) and strong (h) cooling regimes, with the photons being emitted by electrons and examined in the local $\mathbf{E}\times \mathbf{B}$ frame. All photons are calculated in (a)-(d), while the photons emitted during the time interval $\delta t \in [380\pi/\omega_{pe}, 400\pi/\omega_{pe}]$ are chosen  for (e)-(h).}
\end{figure}

The electron pitch-angle distribution fundamentally dictates the emission characteristic of photons through the angle-dependent synchrotron radiation power, $P_{\mathrm{syn}}\propto B_{\mathrm{loc}}^2\gamma _{e}^{2}\sin ^2\theta $.
By substitute the characteristic photon energy, $\varepsilon_{ph}=\hbar \omega_c= \varepsilon_0 \gamma_{e}^{2}\sin \theta $ into the electron scaling laws Eqs.\ (\ref{equation2}) and (\ref{equation3}), we derive the energy-dependence pitch-angle scalings for photons:
\begin{equation}\label{equation7}
  \sin \theta_{ph^<}  \simeq \mathcal{C}^2 \varepsilon_{ph}^{-1}\varepsilon_0
\end{equation}
and 
\begin{equation}\label{equation8}
  \sin \theta_{ph^>}\propto \varepsilon_{ph}^{5/17}\varepsilon_0^{-5/17}\rho_0^{12/17}\mathcal{L}^{-8/17} d_{e0}^{-4/17},
\end{equation}
which are applicable for the low- and high-energy photon scenarios, respectively. Here $\varepsilon_0 = \frac{3}{2}\hbar \omega_L$ and $\omega _L=\left| e \right|B_{\mathrm{loc}}/m_e c$.
While Eq.~(\ref{equation8}) accurately describes the weak-cooling regime, it requires correction for the high-energy tail ($\gamma_e \gtrsim 100$) where radiative effects suppress the electron scaling index to $q < 5/6$. Adopting a corrected index $q \sim 1/6$ based on our simulation results, the photon scaling of Eq.~(\ref{equation8}) in the high-energy limit is modified as
\begin{equation}\label{equation9}
  \sin \theta_{ph^>}\propto \varepsilon_{ph}^{1/13}\varepsilon_0^{-1/13}\rho_0^{12/13}\mathcal{L}^{-8/13}d_{e0}^{-4/13}.
\end{equation}
The intersection of Eqs.~(\ref{equation7}) and (\ref{equation9}) yields the turning-point energy for photons:
\begin{equation}\label{equation10}
  \varepsilon_{t} \propto \mathcal{L}^{4/7}d_{e0}^{2/7}\rho_0^{-6/7}\mathcal{C}^{13/7}\varepsilon_0 \propto \mathcal{L}^{4/7}B_\mathrm{loc}^{13/7}.
\end{equation}
In the strong-cooling scenario, the low-energy photon scaling is governed by,
\begin{equation}\label{equation11}
  \sin \theta_{ph_{\mathrm{cool}}} \simeq \varepsilon_{ph}^{-1/3}\varepsilon_0^{1/3} \mathcal{B}^{4/3} B_{\mathrm{loc}}^{-4/3},
\end{equation}
where $\mathcal{B} = B_{S}\left( m_ec/2\eta t \right)^{1/2}$~\cite{Supplement}.
Correspondingly, the turning-point energy scales as
\begin{equation}\label{equation12}
  \varepsilon_{t,\mathrm{cool}} \propto \mathcal{L}^{3/2}\varepsilon_0 d_{e0}^{3/4}\rho_0^{-9/4}B_\mathrm{loc}^{-13/4} \mathcal{B}^{13/4} \propto \mathcal{L}^{3/2},
\end{equation}
becoming independent of the local magnetic field strength.
These analytical predictions are confirmed by our radiative PIC simulations [Figs.~\ref{fig:figure4}(a) and (b)]. In the hard X-ray regime, the photon pitch angle follows the $ \sin \theta_{ph^{<}}  \sim \varepsilon_{ph} ^{-1}$ scaling. As the energy transitions from hard X-rays to gamma-rays, the scaling index shifts to $r = -d\log \left< \sin \theta_{ph} \right> /d\log \varepsilon_{ph} \sim 1/13$, aligning with Eq. (\ref{equation9}).
Furthermore, parameter scans [Figs. \ref{fig:figure4}(c) and (d)] validate that the photon turning-point energy $\varepsilon_t$ is determined by both $\mathcal{L}$ and $B_{\mathrm{loc}}$ in weak-cooling cases, but is exclusively a function of the energy-injection scale $\mathcal{L}$ under strong cooling.

Figures \ref{fig:figure4}(e) and (g) reveal that the primary acceleration of high-energy electrons occurs in the immediate vicinity of turbulent current sheets. Consequently, the emitted photons reach maximum energies of $\sim 4 m_ec^2$ and $\sim 30 m_ec^2$ at $t=400\pi/\omega_{pe} $ for weak- and strong-cooling regimes, respectively. 
We observe a pronounced statistical correlation between these high-energy photons and turbulent vortex structures. Specifically, photons within these eddies and current sheets possess significantly higher energies compared to those in the ambient region.
To further characterize the spatial dependence of photon anisotropy, we trace the trajectories of a representative sample of $\sim 10^5$ photons during the interval $\delta t \in \left[ 380\pi/\omega_{pe}, 400\pi/\omega_{pe} \right] $ [Figs.\ \ref{fig:figure4}(f) and (h)]. 
In the weak-cooling regime, photons near the eddies exhibit a bimodal angular distribution peaking at $\theta_{ph} = 0 ^{\circ}$ and $180^{\circ}$, whereas those outside the eddies preferentially align toward $\theta_{ph}\simeq90^{\circ}$. 
In contrast, under strong cooling, photons exhibit a dominant trend of moving along magnetic field lines ($\theta_{ph} = 0 ^{\circ}$ or $180^{\circ}$) regardless of their proximity to the eddies. This behavior is a direct consequence of the anisotropy inherited from their parent electrons [see Fig.\ \ref{fig:figure2}(b)].

In summary, we have systematically investigate the interplay between particle pitch-angle anisotropy and synchrotron cooling dynamics in the WEAT regime through a combination of analytical modeling and radiative kinetic simulations. 
We establish a robust scaling model combining the perpendicular drift motions and radiation reactions, capable of predicting the energy-dependent pitch angle and spectral indices across both weak- and strong-cooling regimes. 
Our model not only provides a theoretical foundation for understanding previous numerical observations in both WEAT~\cite{Nattila2022,Vega2025} and large-amplitude turbulence~\cite{ComissoAPJ2019,ComissoAPJL2020,Comisso2021,ComissoAPJL2022},
but also offer a quantitative bridge between microscopic kinetic processes and macroscopic astrophysical observables.
The identified coupling between pitch-angle anisotropy and spectral hardening carries significant implications for extreme astrophysical environments.
Specifically, the pitch-angle-regulated hardening of the lepton energy spectrum offers a natural explanation for the hard radio spectra observed in PWNe~\cite{Lyutikov2019,ComissoAPJL2020}, which require a nonthermal spectral index as hard as $p \lesssim 1.6$.
In addition, the intrinsic anisotropy of the emission process ensures that fast temporal variabilities are preserved rather than smeared out, potentially accounting for the rapid rise and decay times observed in GRB X-ray afterglow flares ~\cite{Beloborodov2011}.
Moreover, our findings suggest that the pronounced electron anisotropy can effectively mitigate the internal cancellation of circular polarization, providing a robust mechanism for the high degrees of circular polarization detected in the optical afterglows of GRBs~\cite{Wiersema2014,Nava2015}.
This work thus establishes a new quantitative paradigm for interpreting multi-messenger signatures from magnetized, radiation-dominated turbulent sources.

This work is supported by the Strategic Priority Research Program of the Chinese Academy of Sciences (Grant No. XDB1550100), the CAS Project for Young Scientists in Basic Research (Grant No. YSBR141), the National Key R\&D Program of China (2025YFF0515103), the National Natural Science Foundation of China (NNSFC) (Grants No. 12447101, No. 12605407, and No. 12675316), and the Postdoctoral Fellowship Program of CPSF (Grant No. GZC20252223). We acknowledge the HPC Cluster of ITP-CAS for providing computational resources. 
The EPOCH code used in this work was in part funded by the UK EPSRC grants EP/G054950/1, EP/G056803/1, EP/G055165/1, EP/M022463/1, and EP/P02212X/1.
We would like to thank Luca Comisso, Muni Zhou, and Kexun Shen for the fruitful discussion.

\bibliography{Reference0604}

\end{document}